\documentclass[aps, prd, twocolumn, amsmath, floats,floatfix, superscriptaddress, nofootinbib,
showpacs]{revtex4}

\usepackage{epsfig}
\usepackage{amsmath,amsfonts,amssymb}
\usepackage[caption=false]{subfig}
\usepackage{verbatim}
\usepackage{float}
\usepackage{morefloats}
\usepackage[dvipsnames]{xcolor}
\usepackage{booktabs}
\usepackage[normalem]{ulem}
\usepackage{multirow}
\usepackage{makecell}
\usepackage{tabulary}
\usepackage{cancel}
\usepackage[T1]{fontenc}

\usepackage[capitalise]{cleveref}

\usepackage{comment}

\crefname{section}{Sec.}{Secs.}
\crefname{table}{Tab.}{Tabs.}
\crefname{figure}{Fig.}{Figs.}
\crefname{equation}{Eq.}{Eqs.}
\crefname{appendix}{Appendix\ }{Appendix\ }

\providecommand{\openone}{\leavevmode\hbox{\small1\kern-3.8pt\normalsize1}}

\newcommand{\s}{\text{s}\,}

\newcommand{\chieff}{$\chi_\mathrm{eff}$}

\definecolor{bostonuniversityred}{rgb}{0.8, 0.0, 0.0}

\usepackage{booktabs}
\usepackage[Q=yes,pverb-linebreak=no]{examplep}
\usepackage{aas_macros}

\begin{document}

\title{\boldmath Exploring gravitational-wave detection and parameter inference \\using Deep Learning methods\unboldmath}

\author{Jo\~ao~D.~\'Alvares}
\email{joaodinis01@gmail.com}
\affiliation{Centro de F\'{\i}sica das Universidades do Minho e do Porto (CF-UM-UP), Universidade do Minho, 4710-057 Braga, Portugal}

\author{Jos\'e~A. Font}
\email{j.antonio.font@uv.es}
\affiliation{Departamento de
  Astronom\'{\i}a y Astrof\'{\i}sica, Universitat de Val\`encia,
  Dr. Moliner 50, 46100, Burjassot (Val\`encia), Spain}
\affiliation{Observatori Astron\`omic, Universitat de Val\`encia,  Catedr\'atico 
  Jos\'e Beltr\'an 2, 46980, Paterna (Val\`encia), Spain}
\author{Felipe~F.~Freitas}
\email{felipefreitas@ua.pt}
\affiliation{Departamento de F\'\i sica da Universidade 
de Aveiro and \\
Centre  for  Research  and  Development  in  Mathematics  and  Applications  (CIDMA)\\
Campus de Santiago, 3810-183 Aveiro, Portugal}

\author{Osvaldo~G.~Freitas}
\email{ogf1996@gmail.com}
\affiliation{Centro de F\'{\i}sica das Universidades do Minho e do Porto (CF-UM-UP), Universidade do Minho, 4710-057 Braga, Portugal}

\author{Ant{\'o}nio~P.~Morais}
\email{aapmorais@ua.pt}
\affiliation{Departamento de F\'\i sica da Universidade 
de Aveiro and \\
Centre  for  Research  and  Development  in  Mathematics  and  Applications  (CIDMA)\\
Campus de Santiago, 3810-183 Aveiro, Portugal}

\author{Solange~Nunes}
\email{solangesilnunes@gmail.com}
\affiliation{Centro de F\'{\i}sica das Universidades do Minho e do Porto (CF-UM-UP), Universidade do Minho, 4710-057 Braga, Portugal}
\email{solangesilnunes@gmail.com}

\author{Antonio~Onofre}
\email{antonio.onofre@cern.ch}
\affiliation{Centro de F\'{\i}sica das Universidades do Minho e do Porto (CF-UM-UP), Universidade do Minho, 4710-057 Braga, Portugal}

\author{Alejandro \surname{Torres-Forn\'e}}
\email{alejandro.torres-forne@aei.mpg.de}
\affiliation{Max Planck Institute for Gravitational Physics (Albert Einstein Institute), D-14476 Potsdam-Golm, Germany}
\affiliation{Departamento de
  Astronom\'{\i}a y Astrof\'{\i}sica, Universitat de Val\`encia,
  Dr. Moliner 50, 46100, Burjassot (Val\`encia), Spain}

\begin{abstract}
We explore machine learning methods to detect gravitational waves (GW) from binary black hole (BBH) mergers using deep learning (DL) algorithms. The DL networks are trained with gravitational waveforms obtained from BBH mergers with component masses randomly sampled in the range from 5 to 100 solar masses and luminosity distances from 100~Mpc to, at least, 2000~Mpc. The GW signal waveforms are injected in public data from the O2 run of the Advanced LIGO and Advanced Virgo detectors, in time windows that do not coincide with those of known detected signals. We demonstrate that DL algorithms, trained with GW signal waveforms at distances of 2000~Mpc, still show high accuracy when detecting closer signals, within the ranges considered in our analysis. Moreover, by combining the results of the three-detector network in a unique RGB image, the single detector performance is improved by as much as 70\%. Furthermore, we train a regression network to perform parameter inference on BBH spectrogram data and apply this network to the events from the the GWTC-1 and GWTC-2 catalogs. Without significant optimization of our algorithms we obtain results that are mostly consistent with published results by the LIGO-Virgo Collaboration. In particular, our predictions for the chirp mass are compatible (up to $3\sigma$) with the official values for 90\% of events.
\end{abstract}

\maketitle

\section{Introduction}
The detection of gravitational waves (GW) from binary black hole (BBH) mergers~\cite{GW150914-prl,GW151226-prl} during the first data-taking run (O1) of Advanced LIGO~\cite{AdvLIGO} was a remarkable milestone that opened up a new window for observing the cosmos. The European detector Advanced Virgo~\cite{Virgo} joined the efforts during the second observing run (O2) which helped improve the sky localization of the sources. Notably, O2 included the first observation of GW from a binary neutron star (BNS) merger, GW170817. This event was accompanied by  electromagnetic radiation which was observed by dozens of telescopes worldwide and brought forth the field of Multi-Messenger Astronomy~\cite{MMA}. During O1 and O2 the LIGO Scientific Collaboration and the Virgo Collaboration (LVC) announced the confident detection of eleven GW signals from compact binary coalescences (CBC)~\cite{GWTC-1}. The third science run (O3) ended on March 2020 after completing almost one full year of data-taking. O3 provided 
a record number of detections, publicly released as low-latency alerts through the GW Catalog Event Database\footnote{gracedb.ligo.org}.
Recently, the LVC has released their second GW transient catalog comprising the 39 CBC detections accomplished in the first six months of O3~\cite{GWTC-2}.

The detection of GW signals from CBC relies on accurate waveform templates against which to perform match-filtered searches. Faithful templates can be built either by solving the gravitational field equations with numerical relativity techniques or by using approximations to the two-body problem in general relativity. Current gravitational waveform models (or approximants) combine analytical and numerical approaches and they are able to describe the entire inspiral-merger-ringdown signal for a large variety of possible configurations of the parameter space (see e.g.~\cite{approx1,approx2,approx3,approx4} and references therein). Once a CBC source is detected, the estimation of its characteristic physical parameters such as component masses, individual spins or distance, is based on Bayesian inference~\cite{LALInference,Bilby}. However, Bayesian inference can be computationally expensive as it may  take of the order of days to obtain sufficient number of posterior samples for BBH~\cite{Green:2020}. The situation aggravates as the number of detections increases, as it is expected in the forthcoming observational campaigns of the GW detector network. As an example, the predicted detection count of BBH mergers in one-calendar-year  observing run of the network during O4 is $79^{+89}_{-44}$~\cite{prospects}.  To overcome this difficulty Deep Learning (DL) algorithms constitute an attractive choice 
to speed-up parameter estimation \cite{Gabbard:2018,Gabbard:2019,Green:2020,Wang:2020,Chua:2020}.

Machine learning (ML) and DL are bringing about a revolution in data analysis across a variety of fields and GW astronomy is not alien to that trend. In particular, the use of Deep Neural Networks (DNN) for classification and/or prediction tasks has become the standard on data analysis applications, ranging from industrial applications \cite{RUMMEL1989203,JAIN1989283,DOM1989257} and medical diagnosis \cite{MAMMONE1989185,JMAI5545} to particle physics \cite{Alves:2019ppy,Freitas:2019hbk,Csaki:2018hyw} and cosmology \cite{DBLP:journals/corr/abs-1904-05146,Hong:2019qix}.
This trend has now  organically been extended to  GW astronomy, both for signal detection \cite{Gebhard:2019ldz,Lin:2020aps,Sadeh:2020txb} and for detector characterization, by reducing the impact of noise artifacts or ``glitches'' of instrumental and environmental origin~\cite{Biswas:2013,Powell:2015,Powell:2017,Razzano:2018,Cavaglia:2018,George:2018,Miquel:2019,Coughlin:2019, Colgan:2020}. 
Glitches can potentially affect GW detection as they contribute to the background in transient searches, decreasing the statistical significance and increasing the false alarm ratio of actual GW events. For this reason, the detection and classification of glitches has become an important application of DL in GW astronomy. Worth highlighting is the Gravity Spy project \cite{GravitySpy} which combines DL with citizen science to identify and label families of glitches from the twin Advanced LIGO detectors. Recent approaches to eliminate, or at least mitigate, the effect of glitches are discussed in \cite{Driggers:2019,Vajente:2020,Torres:2020}. 
We also note that recent developments of GW signal denoising algorithms include Total-Variation methods ~\cite{Torres:2014,Torres-Forne:2018yvv}, Dictionary Learning approaches \cite{Torres:2016}, DL autoencoders and Deep Filtering \cite{Shen:2019} or Wavenets \cite{Wei:2020}. 

Contrary to CBC signals there are other types of GW sources whose detection is not template-based, namely burst sources,  continuos-wave sources and stochastic sources. Archetypal examples of the first two are 
core-collapse supernovae (CCSN) and rotating neutron stars (RNS), respectively. In the case of CCSN the number of waveform templates available is fairly scarce due to the complexity of the very computationally expensive numerical simulations required to model the supernova mechanism, rendering a comprehensive survey of the vast parameter space of the problem impracticable. In the case of RNS, their continous, monochromatic GW signals are very stable, which makes match-filtering-based techniques unfeasible due to the computational resources the task would involve. Both sources are however excellent candidates for DL methods and specific pipelines have been developed in the two cases. Recent approaches that use DL to improve the chances of detection of CCSN GW signals are discussed in~\cite{Astone:2018,Chan:2019,Cavaglia:2020}. Additionally, as DL methods are designed to efficiently deal with large amounts of data, they could offer a very effective solution to detect and analyze signals from 
RNS~\cite{Miller:2019,Beheshtipour:2020,Morawski:2020,Bayley:2020}.

In this paper we explore the use of computer vision techniques and DL methods to both detect GW from BBH mergers and perform parameter inference using RGB spectrograms that combine open data from the Advanced LIGO and Advanced Virgo three-detector network. To achieve our goal we train a cross-residual network which allows us to extract information about source parameters such as luminosity distance, chirp mass, network antenna power, and effective spin. As we show below, the application of our residual network to the BBH detections included in the GWTC-1 and GWTC-2 catalogs yields a remarkable agreement with the LVC results.

This paper is organized as follows:  Section \ref{sec:GW_generation}  describes the generation of the GW datasets used for training and testing. Section \ref{sec:DL_network} deals with the general characteristics of our Deep Neural Network, describing its various architectures and the  methodology employed for classification and regression. 
The assessment of the network is discussed in Section \ref{sec:network_assessment} and our main results regarding the use of our network for the analysis (detection and inference) of real GW events is presented in Section \ref{sec:comparison}.
Finally, our main conclusions are summarized in Section \ref{sec:conclusions}.  


\section{GW Datasets Generation \label{sec:GW_generation}}

We begin by describing the generation of the datasets used in our analysis. All CBC waveforms employed in the classification datasets were obtained using pyCBC \cite{alex_nitz_2020_4075326} with the {\tt SEOBNRv4\_ROM} approximant while the regression datasets use {\tt SEOBNRv4HM\_ROM} \cite{Cotesta:2020qhw}, {\tt IMRPhenomPv2} \cite{PhysRevLett.113.151101} and {\tt IMRPhenomD} \cite{PhysRevD.93.044006, PhysRevD.95.044028}.
For the sake of simplicity we start by considering spinless black holes and quasi-circular binaries with no orbital eccentricity. Furthermore, since current GW detector networks are far more sensitive to the plus polarization than to the cross one, we only generate plus-polarized waves. We note that this has the drawback of making impossible to break the degeneracy between luminosity distance and inclination.

\vspace*{3mm}
\noindent
\textit{Single detector waveforms for classification}
\vspace*{1.5mm}

The purpose of this first dataset is to allow our DL models to discern the presence of a GW signal with data collected from a single detector. In particular we employ a 500~s noise segment from the Hanford detector with initial GPS time $t_\mathrm{GPS} = 1187058342~\s$. Defining $\tau$ to be some time from the start of our noise segment, we randomly select $\tau_0\in[5,495]\,\s$ and isolate the window $[\tau_0-5\,\s,\tau_0+5\,\s]$. This strain selection, which we denote as $n$, is then whitened through inverse spectrum truncation, using its own amplitude spectral density (ASD). Then, we apply a bandpass filter from 20~Hz to 300~Hz, as well as notch filters at the individual frequencies 60~Hz, 120~Hz and 240~Hz.
For the generation of the waveform signal strain $h$, a random pair of black hole masses $(m_1,m_2) \in U\left([5,100]\right)~M_\odot$ is selected for a BBH merger with luminosity distance $d_L=2000~\mathrm{Mpc}$ and inclination $\iota=\frac{\pi}{2}$. This waveform is whitened using the ASD of the selected noise strain $n$ and the same filtering process is undertaken. The resulting waveform is injected into the noise window in such a way that the maximum amplitude occurs at $\tau_0$. Following this, the constant-Q transform is calculated for the $[\tau_0-0.16\,\s, \tau_0+0.4\,\s]$ interval in the composite signal $S=h+t$, and a spectrogram is produced. A second spectrogram without signal injection is also generated for the same interval. Both spectrograms are saved as images and appropriately labeled as ``signal'' and ``background''. This process is iterated 5000 times to build our dataset. The same procedure is taken for the luminosity distances $d_L = 100,~300,~1000,~1500~\text{and}~2000~\mathrm{Mpc}$. A summary of the single detector classification procedure is shown in the first row of Table~\ref{tab:datasets}.

\begin{figure}[t]
\begin{center}

\subfloat[]{%
\includegraphics[clip,scale=0.86]{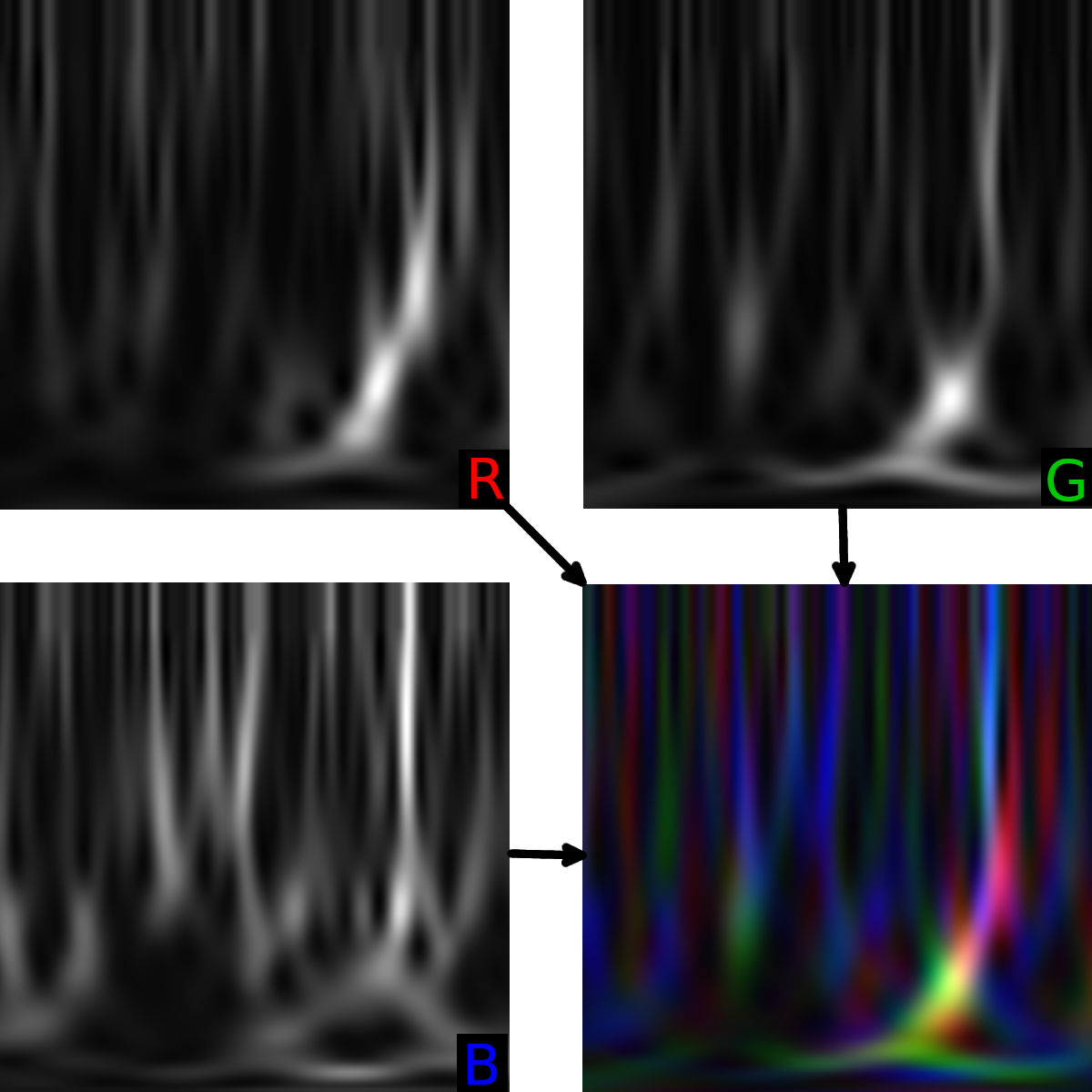}%
}
\\
\subfloat[]{%
\hspace*{-1mm}\includegraphics[clip,scale=1.9]{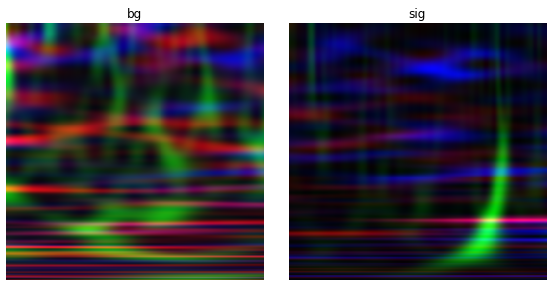}%
}

\caption{{\bf (a)} Combining single detector spectrogram data into a single RGB image, to be used combined by the deep learning network architectures. The Hanford (top-left), Livingston (top-right) and Virgo (bottom-left) spectrogram data, are used as the Red, Green and Blue images, respectively, to build the full RGB image (bottom-right).
{\bf (b)} RGB image from background labeled spectrogram (left) as compared with a spectrogram where a GW waveform was injected into real conditions noise(right).}
\label{fig:RGBimage}
\end{center}
\end{figure}

\vspace*{3mm}
\noindent
\textit{Multiple detector waveforms for classification}
\vspace*{1.5mm}

In order to combine the data from all three detectors (Hanford, Livingston and Virgo) we select coincident segments of  $500\,\s$ from all detectors  starting at a certain $t_\mathrm{GPS}$ time. The process is then identical to that of a single detector case with $\tau_0$, $m_1$ and $m_2$ randomly generated. A $[\tau_0-5\s,\tau_0+5\s]$ time window is extracted from the longer segments for the three  detectors. The resulting background strain data, $n_H$, $n_L$ and $n_V$, for Hanford, Livingston and Virgo interferometers respectively, is treated in the same way as described above. However, when injecting a signal, one must make sure that the specific ASD of each detector is being used. After the generation of the signal waveform and its injection into the background noise segments, we include the antenna  power from each detector into our time series $S_H$, $S_L$ and $S_V$. At this stage, we can emulate the sky position for the signal by randomly choosing one of three detectors as a reference, and shift the beginning of the other two time series according to their time delay with respect to the reference detector. 
Once the three spectrograms are produced they are combined into a $560\times560\times3$ array in such a way that each of them is represented by a certain colour channel in a RGB image. Specifically, Hanford, Livingston and Virgo datasets are mapped into the Red, Green and Blue channels respectively, as can be seen in \cref{fig:RGBimage}(a). 
As in the single detector case, an equivalent background spectrogram without signal injection is produced. Both, background (left) and signal (right) spectrograms, are represented in \cref{fig:RGBimage}(b). Once again, this process is iterated 5000 times for each luminosity distance we consider. The classification procedure for the multiple detector case is summarized in the second row of Table~\ref{tab:datasets}.

\vspace*{3mm}
\noindent
\textit{Mass dependent dataset}
\vspace*{1.5mm} 

The purpose of this dataset is to check how the trained models perform depending on the binary component masses and luminosity distances, both for the single and multiple detector cases. To this end, we consider a number of mass combinations, where for each $m_1$, 
ranging from 5 to 100~$M_\odot$ in steps of 2~$M_\odot$, there is a set of values for $m_2$, in the range [$m_1$,100]~$M_\odot$, covered also with steps of 2~$M_\odot$. Regarding the luminosity distance $d_L$, we use distances ranging from 100 to 2000~Mpc, in steps of 100~Mpc. The inclination is kept fixed at $\iota=\frac{\pi}{2}$. For each $d_L$ and each of the 1225 $(m_1,m_2)$ mass combinations, a waveform is generated and injected into the detector's noise following the same procedure as described above. However, in this case, only the spectrograms with the injections are saved and labeled with the corresponding distance and mass.

\vspace*{3mm}
\noindent
\textit{Regression datasets}
\vspace*{1.5mm}

\begin{figure}
\begin{center}
\begin{tabular}{cc}
\hspace*{-5mm}
\epsfig{file=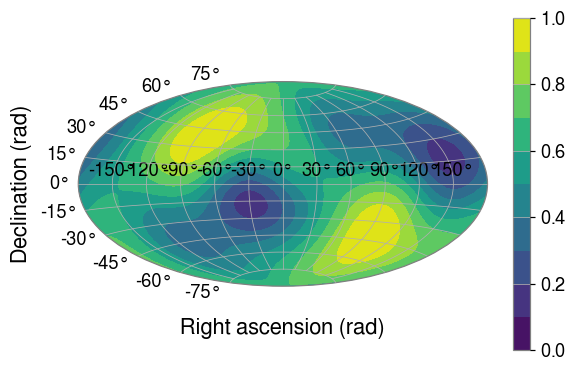,height=5.5cm,clip=} \\
\end{tabular}
\caption{Network Antenna Power as a function of sky position for (t mod 24h)=9742.}
\label{fig:skymap}
\end{center}
\end{figure}

We also study how the DL algorithms can be used to extract information about the physical parameters from the generated data. This procedure is typically denoted as regression. For this purpose, a larger dataset was deemed necessary and only the multiple detector case, was considered. To avoid a dependence on a particular approximant, three different datasets for each of the approximants {\tt SEOBNRv4HM\_ROM}  \cite{Cotesta:2020qhw}, {\tt IMRPhenomPv2} \cite{PhysRevLett.113.151101} and {\tt IMRPhenomD} \cite{PhysRevD.93.044006, PhysRevD.95.044028} were built. It is also relevant to note that, due to an apparent degradation of the regression close to the upper range of the sampled distances, we decided to consider distances up to 4~Gpc, although we set our range of operation to go up to a maximum distance of 2.5~Gpc. Here, we do not build separate datasets for particular values of the distance but instead let $d_L$ be randomly generated within this range. The component masses $m_1$ and $m_2$ are again randomly sampled in the $[5,100]~M_\odot$ interval while the inclination takes a random value in the $[0,\pi]$ interval. We also sample the sky position by taking into account the antenna pattern of each detector. An example of the network antenna power is shown in Fig.~\ref{fig:skymap}. In our regression dataset black holes are assumed to have a dimensionless spin in the range $[-1,1]$ and those are aligned with the orbital angular momentum, allowing to compute the effective inspiral spin, $\chi_\mathrm{eff}$,
\begin{eqnarray}
\chi_{\rm eff}=\frac{m_1\chi_1^{\parallel}+m_2\chi_2^{\parallel}}{m_1+m_2}, 
\end{eqnarray}
where $\chi^{\parallel}_i$ is the component of the $i$-th spin along the orbital angular momentum. Since we assume from the beginning that a given input to the regression model will necessarily contain a GW signal of some sort, we need not to worry about generating the background-only cases. Furthermore we impose a threshold for the signal-to-noise ratio (SNR) so that we only allow cases where $\mathrm{SNR}>5$. The sizes of the {\tt SEOBNRv4HM\_ROM}, {\tt IMRPhenomPv2} and {\tt IMRPhenomD} datasets are, respectively, 15961, 12922, and 13281.
An extra dataset focused on lower masses ($m_1, m_2\in[5,35]M_\odot$) was also generated using the {\tt SEOBNRv4HM\_ROM} approximant, containing 15538 events. This was combined with the original {\tt SEOBNRv4HM\_ROM} dataset for a lower-mass weighted dataset with a total of 31499 items. All this information is summarized in Table  \ref{tab:datasets}.

\section{Deep Neural Network: architectures and methodologies \label{sec:DL_network}}

As mentioned in \cref{sec:GW_generation} we encode the  information of the waveforms produced by BBH mergers into a spectrogram. Here we describe how to apply our DL algorithms to identify GW information from spectrogram data. This is done using the classification and regression networks that we discuss in this section. A summary of the architectures of both networks is reported in Table \ref{tab:archs}.



\begin{table*}[!htbp]
\setlength\doublerulesep{4mm} 
\setlength\tabcolsep{0pt}
\begin{tabular}{|c|c|c|}
\toprule
\hline
\multicolumn{3}{|c|}{Classification } \\
\hline
Parameters & Train size & Validation size \\
\hline
\makecell{Single detector\\ $(m_1,m_2)\sim U(5,100) \ M_\odot$, \\
         $d_L = [100, 300, 1000, 1500, 2000]$ Mpc, \\
         $\iota=\frac{\pi}{2}$, \\
         approximant: {\tt SEOBNRv4\_ROM}} & \makecell{4000 images \\
                                            560 $\times$ 560 pixels ~\\
                                            8-bit gray scale} 
                                & \makecell{1000 images \\
                                            560 $\times$ 560 pixels ~\\
                                            8-bit gray scale} 
                                \\
\hline                                
Total images &  20000 & 5000 \\                      
\hline
\makecell{Multiple detector \\ (same parameters as above)} & \makecell{4000 images \\
                                            560 $\times$ 560 pixels ~\\
                                            8-bit RGB} 
                                & \makecell{1000 images \\
                                            560 $\times$ 560 pixels ~\\
                                            8-bit RGB} 
                                \\      
\hline                                
Total images &  20000 & 5000 \\                                
\midrule
\hline
\hline
 \multicolumn{3}{|c|}{Regression } \\
\hline
Parameters & Train size & Validation size \\
\hline
\makecell{Multiple detector\\ $(m_1,m_2) \sim U(5,100) \ M_\odot$, \\
         $d_L \sim U(100, 4000)$ Mpc, \\
         $\iota \sim U(0,\pi)$, \\
         $spin \sim U(-1, 1)$, \\
         restriction: $\mathrm{SNR>5}$ \\
         approximant: {\tt SEOBNRv4HM\_ROM}} & \makecell{12769 images \\
                                            224 $\times$ 224 pixels ~\\
                                            8-bit RGB} 
                                & \makecell{3192 images \\
                                            224 $\times$ 224 pixels ~\\
                                            8-bit RGB} 
                                \\      
\hline                          
\makecell{(same parameters as above)\\approximant: {\tt IMRPhenomPv2}} & \makecell{ 10338 images \\
                                            224 $\times$ 224 pixels ~\\
                                            8-bit RGB} 
                                & \makecell{ 2584 images \\
                                            224 $\times$ 224 pixels ~\\
                                            8-bit RGB} 
                                \\      
\hline                          
\makecell{(same parameters as above)\\approximant: {\tt IMRPhenomD}} & \makecell{ 10625 images \\
                                            224 $\times$ 224 pixels ~\\
                                            8-bit RGB} 
                                & \makecell{ 2656 images \\
                                            224 $\times$ 224 pixels ~\\
                                            8-bit RGB} 
                                \\    
\hline                                
Total images &  43009 & 14689 \\                                
\hline
\multicolumn{3}{|c|}{Extra dataset} \\                                
\hline
\makecell{Multiple detector\\ $(m_1,m_2) \sim U(5,35) \ M_\odot$, \\
         $d_L \sim U(100, 4000)$ Mpc, \\
         $\iota \sim U(0,\pi)$, \\
         $spin \sim U(-1, 1)$, \\
         restriction: $\mathrm{SNR>5}$ \\
         approximant: {\tt SEOBNRv4HM\_ROM}} & \multicolumn{2}{|c|}{\makecell{15538 images \\
                                            224 $\times$ 224 pixels ~\\
                                            8-bit RGB}}
                                \\    
\midrule                                
\hline
\end{tabular}
\caption{Description of the full classification and regression datasets for training and validation with both single and multiple detectors. The images are generated from the wave-forms calculated by pyCBC. For the classification datasets, the individual masses $(m_1,m_2)$ are sampled with an uniform distribution within the range of 5 to 100 $M_\odot$. For the regression dataset the parameters of individual masses, distances, inclination and spin are also uniformly sampled. The extra dataset is generated in order to include more examples of small masses and complement the dataset generated with the {\tt SEOBNRv4HM\_ROM} approximant.}
\label{tab:datasets}
\end{table*}

\subsection{Classification Network \label{sec:class_net}}

Our first task is to test whether a Deep Convolutional Network can distinguish between possible signal events over a random background. For this we choose a Residual Network (ResNet) as our base architecture. ResNets were first proposed in \cite{HeZRS15} and consist of a DNN built as blocks of convolutional layers together with short cut connections (or skip layers) 
in order to avoid the well known gradient vanishing/exploding problem (see~\cite{279181} for details). In our analysis, we have tested the discriminant power of the ResNet with an increasing number of layers, namely ResNet-18, ResNet-34, ResNet-50, and ResNet-101, using the data set described in~\cref{sec:GW_generation}.

\begin{table*}[!htb]
\setlength\tabcolsep{0pt}
\begin{tabular}{|c|c|c|}
\toprule
\hline
\multicolumn{3}{|c|}{Classification } \\
\hline
Base architecture ~ & Hyperparameters ~ & Metric performanc (AUC, PPV, RMSE) ~\\
\hline
\makecell{ResNet-101 \\ + custom header} & \makecell{input size: $275 \times 275 \times 3 (1)$, \\
                                                 batch size: 8 images, \\
                                                 learning rate: $[0.001,0.1]$, \\
                                                 weight decay: $0.0001$ \\
                                                 loss function: Cross Entropy Loss (CE)} 
                                     & \makecell{Single detector AUC: 0.72 \\ 
                                                 Multiple detecors AUC: 0.82}  \\
                                                 
\midrule                                
\hline
\hline
\multicolumn{3}{|c|}{Regression} \\
\hline
Base architecture ~ & Hyperparameters ~ & RMSE ~\\
\hline
\makecell{xResNet-18 \\
         + Blur average layer \\
         + MC Dropout \\
         + custom header} & \makecell{input size: $128 \times 128 \times 3$, \\
                                                 batch size: 64 images, \\
                                                 max learning rate: $[0.001,0.1]$, \\
                                                 weight decay: $0.0001$, \\
                                                 loss function: Mean Squared Error (MSE)} 
                                     & RMSE: 0.021 \\

\midrule                                

\hline
\end{tabular}
\caption{Convolutional Neural Networks architectures employed for the classification, multiple (single) detectors, and regression tasks. The custom header for the classification CNN is described in ~\cref{custom_head:1}, the custom header for the regression model has the same structure with the main difference that the final layer has only one unit with a linear activation function.}
\label{tab:archs}
\end{table*}

For the classification task the highest discriminant power was achieved with a ResNet-101 (see Table \ref{tab:archs}), which consists of 101 layers, where in between each Conv2D layer we have a series of batch normalizations, average pooling and rectified activations (ReLU). For our task, we have replaced the last fully connected layers of the ResNet-101, responsible for the classification, with the following sequence of layers:

\begin{itemize}
\item \label{custom_head:1} an adaptive concatenate pooling layer ({\tt AdaptiveConcatPool2d}), 
\item \label{custom_head:2} a flatten layer,
\item \label{custom_head:3} a block with batch normalization, dropout, linear, and ReLU layers,
\item \label{custom_head:4} a dense linear layer with 2 units as outputs, each unit corresponding to \textit{signal} or \textit{background} class and a softmax activation function.
\end{itemize}

The {\tt AdaptiveConcatPool2d} layer uses adaptive average pooling and adaptive max pooling and concatenates them both.
One important aspect of the training of DNN models, and yet often not given the due attention, is the choice of the batch size. The use of large batch sizes helps the optimization algorithms avoiding overfitting \cite{DBLP:Samuel,DBLP:Smith15a, Akiba2017ExtremelyLM} acting as a regularizer. However, the batch size is ultimately bounded by the amount of memory available in a given hardware. One way to work around this limitation is the use of mixed precision training~\cite{DBLP:Micikevicius}. This method uses half-precision floating point numbers, without losing model accuracy or having  to modify hyper-parameters. This nearly halves memory requirements and, on recent GPUs, speeds up arithmetic calculations. 

The learning rate and weight decay are other two key hyperparameters to train DNNs. A good choice of these two parameters can greatly improve the model performance. In our particular case it implies a high accuracy classification and a good background rejection, while drastically reducing the training time. Instead of using a fixed value for the learning rate we opted to use the so called Cyclical Learning Rates (CLR)~\cite{DBLP:Smith15a}. To this end one must specify the minimum and maximum learning rate boundaries as well as a step size. While the latter corresponds to the number of iterations used for each step, a cycle consists of two such steps: one in which the learning rate increases and the other in which it decreases~\cite{DBLP:Smith15a}. Following the guidelines from Ref.~\cite{DBLP:1803-09820}, we have performed a scan over a selected range of values for both learning rates and weight decays. According to \cite{DBLP:1803-09820} the best initial values for learning rates are the ones who give the steepest gradient towards the minimum loss value. In our case, we have found it to be $2\times10^{-3}$ for the learning rate and $1\times10^{-5}$ for the weight decay, while for the maximum learning rate value we just multiply the initial value by 10.

\subsection{Regression Network \label{sec:regression_net}}

The off-the-shelf ResNet models used in the classifier did not provide a very good performance when applied to the regression task. As such, an alternative model was sought. We based the regression network architecture on a Cross-Residual Network (xResNet; see Table \ref{tab:archs}) \cite{2018arXiv181201187H} and following the guidelines in \cite{2019arXiv190411486Z} replaced the average pooling layers with blur pooling ones. Furthermore, we made use of Dropout layers before pooling in order to approximate a Bayesian variational inference process. This has given us a way of estimating the network's uncertainty in the parameter inference at testing time using Monte Carlo (MC) dropout \cite{2015arXiv150602158G}. For training, we use once again the CLR, with $1\times10^{-2}$ as the initial value for the learning rate and $1\times10^{-3}$ as the weight decay.
It is important to mention that we use the spectrogram images generated from the GW signals to infer continuous values for variables such as the chirp mass $\mathcal{M}$ of the BBH system 
\begin{eqnarray}
\mathcal{M}\equiv\frac{(m_1 m_2)^{3/5}}{(m_1+m_2)^{1/5}}\,,
\end{eqnarray}
or the luminosity distance of the source $d_L$. While this approach seems to be rather non-intuitive, CNN's carry inductive biases rooted in translation invariances. Such biases are a direct consequence of the convolutional filters and can be used to extract information from patterns in the spectrogram images and correlated to physical continuous variables.

\section{Network Assessment \label{sec:network_assessment}}

\subsection{Classifier}
\subsubsection{Single Detector Performance \label{sec:single_detector}}

\begin{figure}[t]
\centering
    \vspace{-0.4cm}
    \subfloat{%
    \includegraphics[clip,scale=0.29]{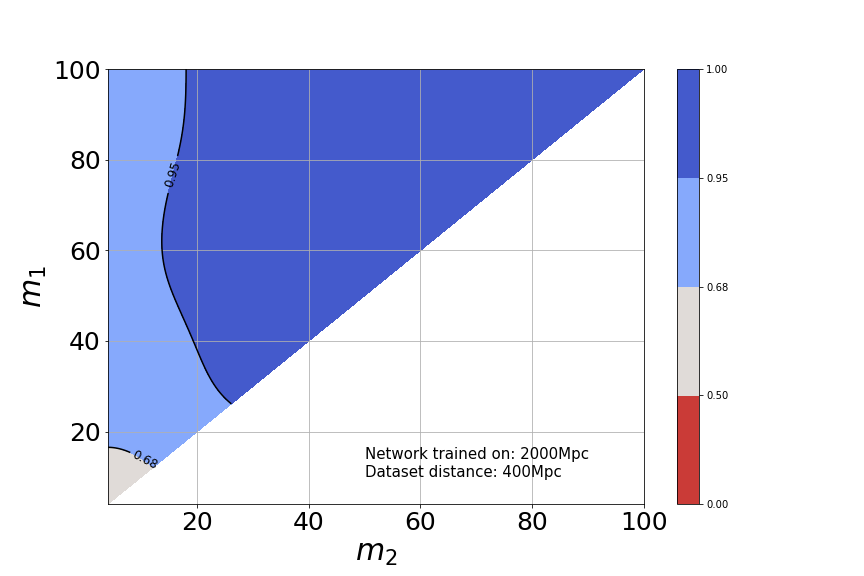}%
    }\\
    \vspace{-0.7cm}
    \subfloat{%
    \includegraphics[clip,scale=0.29]{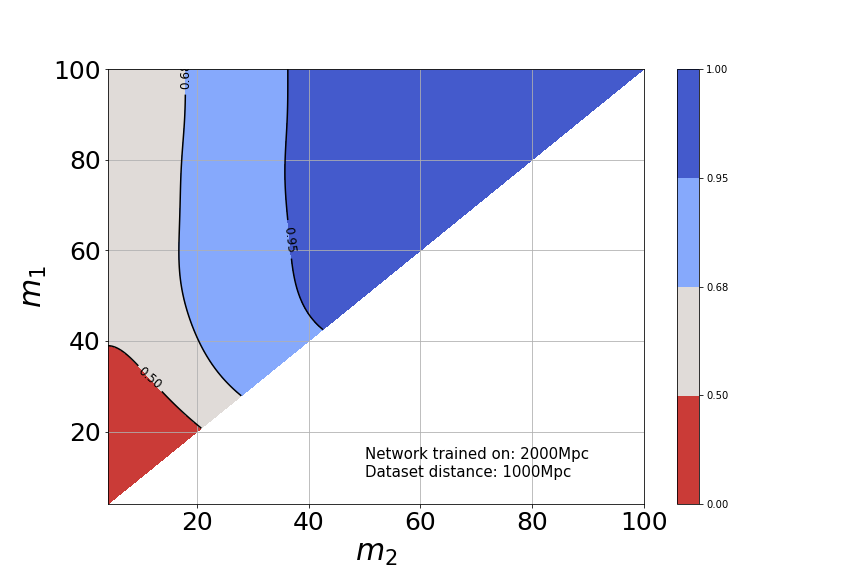}%
    }
    \\
    \vspace{-0.7cm}
    \subfloat{%
    \includegraphics[clip,scale=0.29]{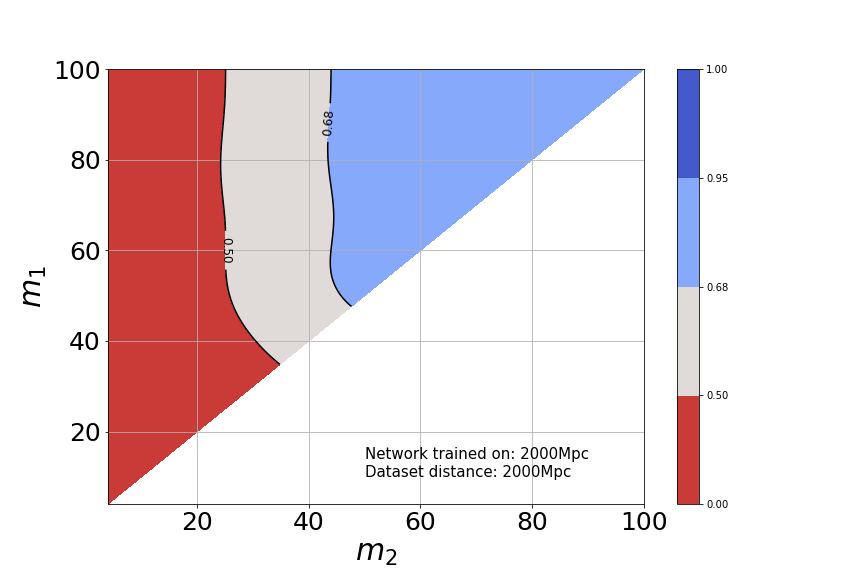}%
    }
    \\    
\caption{Simulated signal scores using one single detector for different luminosity distances and evaluated with DL networks trained with GW waveforms from BBH mergers at a luminosity distance of 2000~Mpc. Results are shown as a function of the BH masses of the binary system, $m_1$ and $m_2$, for GW signals from sources at 400~Mpc (top), 1000~Mpc (middle) and 2000~Mpc (bottom).}
\label{fig:singledetector_scores}
\end{figure}

\begin{figure}[t]

\centering
    \vspace{-0.4cm}
    \subfloat{%
    \includegraphics[clip,scale=0.29]{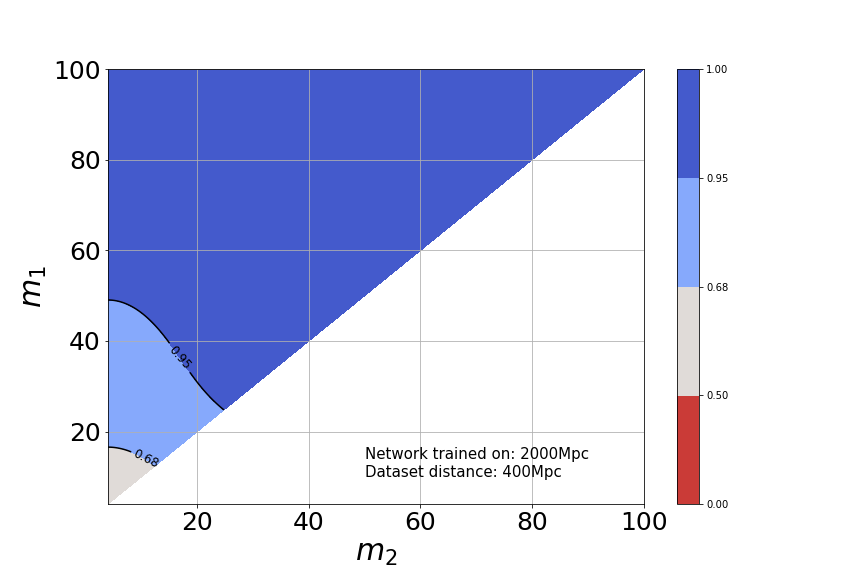}%
    }
    \\
    \vspace{-0.7cm}
    \subfloat{%
    \includegraphics[clip,scale=0.29]{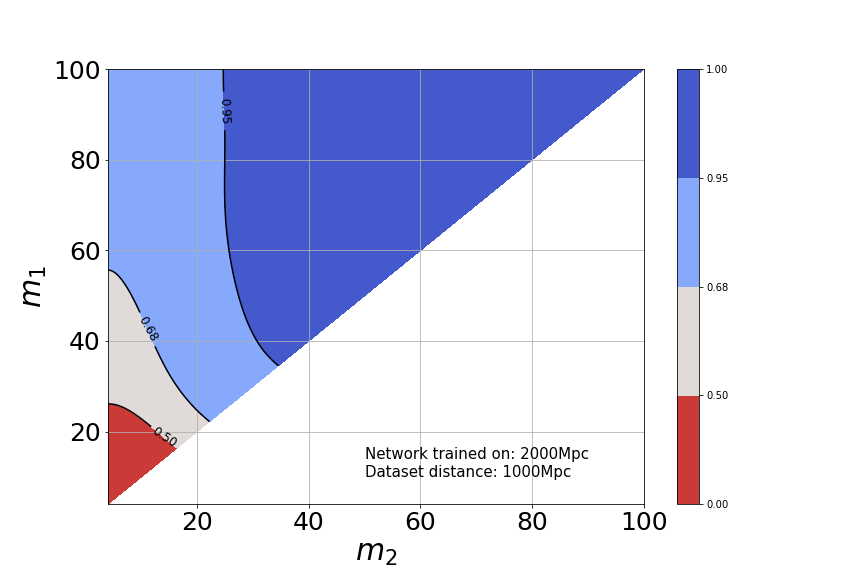}%
    }
    \\
    \vspace{-0.7cm}
    \subfloat{%
    \includegraphics[clip,scale=0.29]{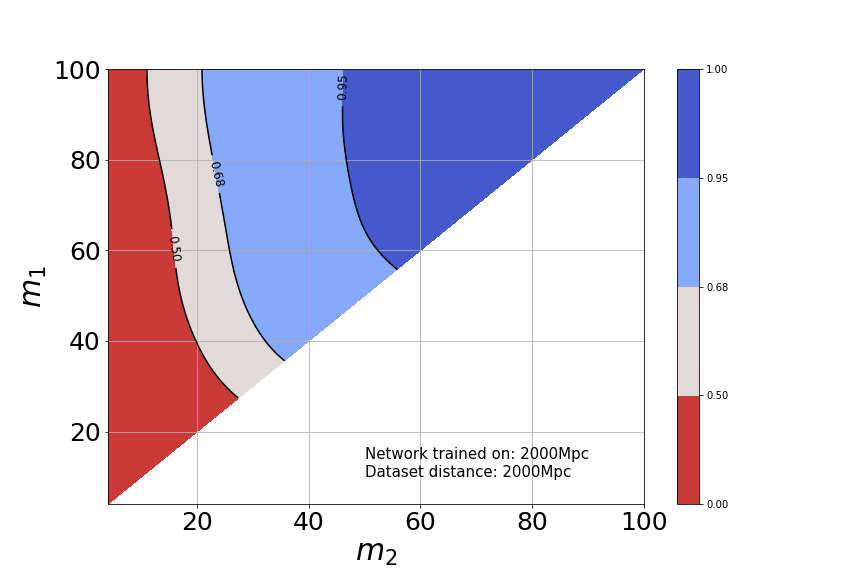}%
    }
    \\    
\caption{Same as Fig.~\ref{fig:singledetector_scores} but employing multiple detectors to estimate the simulated 
signal scores displayed.}
\label{fig:multipledetector_scores}
\end{figure}

For the single detector case we find that the model accuracy increases when trained with GW signals injected at larger distances, up to an accuracy of 72\%, with a positive predictive value (PPV) of 80\%, for the network trained on signals injected a 2~Gpc. In \cref{fig:singledetector_scores} we show contour plots of the scores given to the events in the mass dependence dataset for this last, top-performing model. It is evident from the figure that these scores are higher for lower luminosity distances and higher component masses. This trend is not wholly surprising given that the waveform amplitude is inversely proportional to $d_L$ and proportional to the chirp mass $\mathcal{M}$, and therefore such signals should be expected to have a higher SNR. It is worth stressing that the scores increase for GW signals from sources at shorter distances, even when the network is trained with signals at 2~Gpc. This may suggest that, when searching for potential GW signals from BBH mergers with DL networks, the training phase can start by using signals from far away sources. The low mass region in the ($m_1$,$m_2$) plane remains, however, an issue, particularly as the dataset distance increases, with scores below 0.5. A dedicated study of this issue will be presented elsewhere.

\subsubsection{Multiple Detector Performance \label{sec:multiple_detector}}
\begin{figure}[t]

\centering
    \subfloat{%
    \hspace*{-7mm}\includegraphics[clip,scale=0.36]{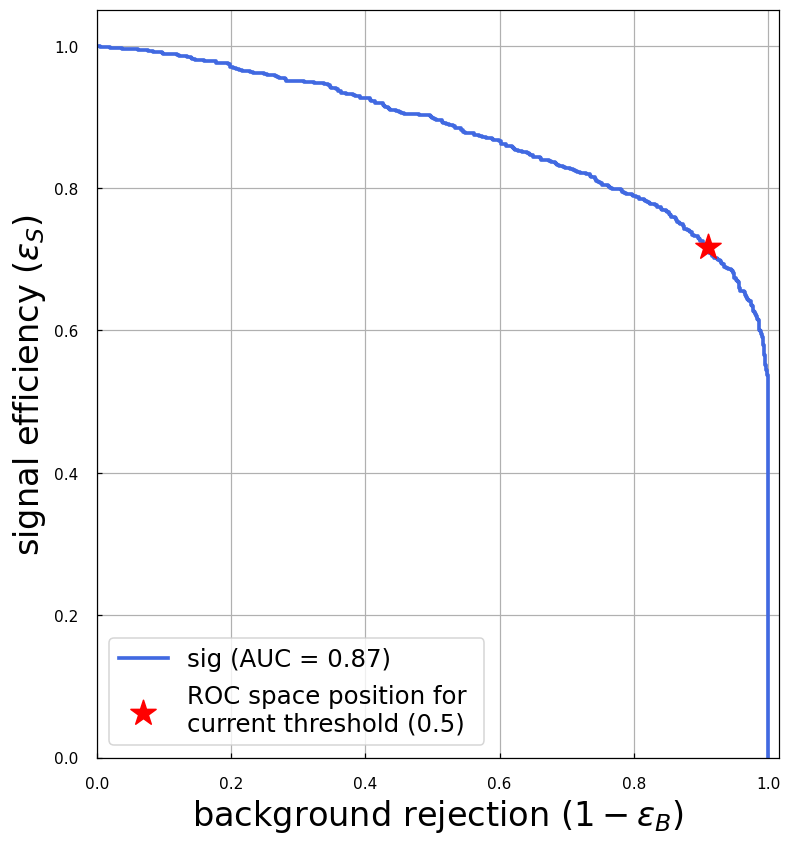}%
    }

\caption{ROC curve for the best-performing classifier. The red star displays the current threshold location on the ROC used for classify the events into signal ($\text{score} \geq 0.5$) or background ($\text{score} \leq 0.5$).}
\label{fig:ROC}

\end{figure}

For the multiple detector case we observe the same trend as in the single detector scenario. However, the performance significantly improves in what concerns the distance of the sources used in the training set. Combining three detectors yields an across-the-board improvement in accuracy of up to 82\% (90\% PPV). \cref{fig:multipledetector_scores} shows the performance of the best multiple detector model as a function of the binary component masses. Once again, it is noticeable that larger masses and smaller distances result in higher scores. Comparing \cref{fig:multipledetector_scores} and  \cref{fig:singledetector_scores} shows that the  multiple detector model yields more confident results, as the region of $\mathrm{score}>0.95$ (dark blue) is overall more prominent. \cref{fig:ROC} exhibits the receiver operating characteristic (ROC) curve for our best-performing network, using 2000Mpc data. The $x$-axis shows the fraction of background-only spectrograms that are successfully rejected by the network, $1-\varepsilon_B$, while the $y$-axis represents the fraction of successfully detected signals, $\varepsilon_S$. As can be seen in the ROC curve, we could alter the threshold for classification to be more or less strict, according to the necessities of the problem at hand.

\subsection{Regression}

\subsubsection{Luminosity Distance Regression \label{sec:distance_regression}}

\begin{figure}[t]

\centering
    \subfloat{%
    \hspace*{-7mm}\includegraphics[clip,scale=0.41]{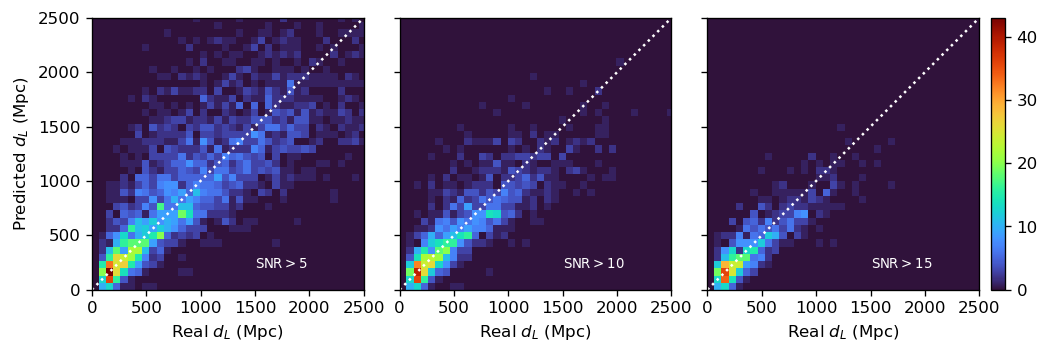}%
    }
    \\
    \subfloat{%
    \hspace*{-7mm}\includegraphics[clip,scale=0.41]{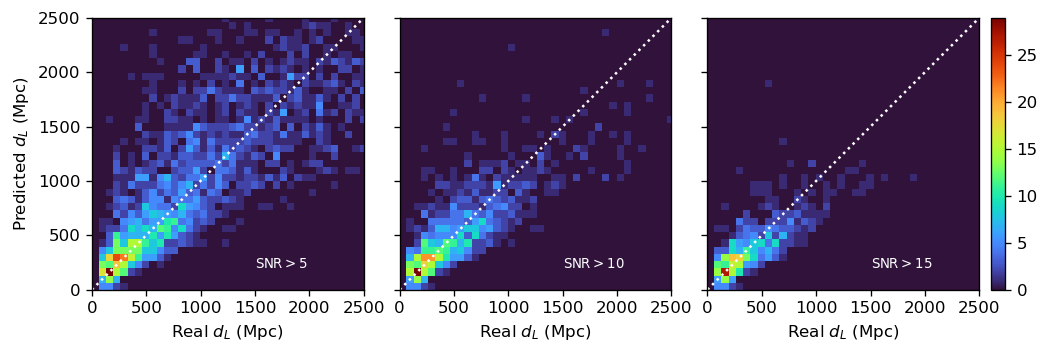}%
    }
    \\
    \subfloat{%
    \hspace*{-7mm}\includegraphics[clip,scale=0.41]{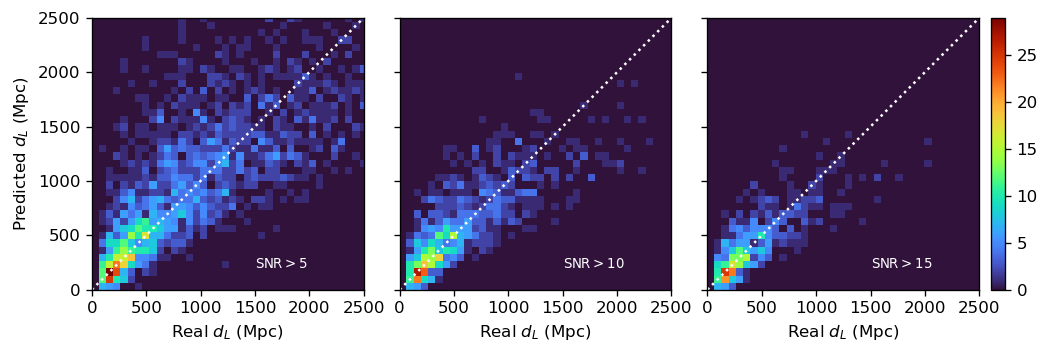}%
    }
    \\
\caption{Calibration results for $d_L$ using the different approximant datasets, {\tt SEOBNRv4HM\_ROM} (top), {\tt IMRPhenomPv2} (middle) and {\tt IMRPhenomD} (bottom), and for different SNR thresholds.}
\label{fig:multipleapprox_dL}

\end{figure}

\cref{fig:multipleapprox_dL} shows the performance of $d_L$ regression for each trained model on its respective validation set. Each event is evaluated 100 times using MC dropout and the mean value of the regressed parameter outputs are stored in the histograms. The white dashed diagonal line shows the ideal behaviour. For the lowest SNR threshold (left columns) the deviation from the ideal behaviour can be quite large. However, as the threshold increases to SNR>10 and SNR>15 we are able to more confidently resolve the distances. On the other hand, these thresholds mean that we lose the ability to resolve larger distances for which larger SNRs are also rarer. All three  approximants exhibit roughly the same behaviour with the {\tt SEOBNRv4HM\_ROM} dataset showing a slightly tighter distribution.

\subsubsection{Network Antenna Power Regression \label{sec:NAP_regression}}

\begin{figure}[t]

\centering
    \subfloat{%
    \hspace*{-7mm}\includegraphics[clip,scale=0.41]{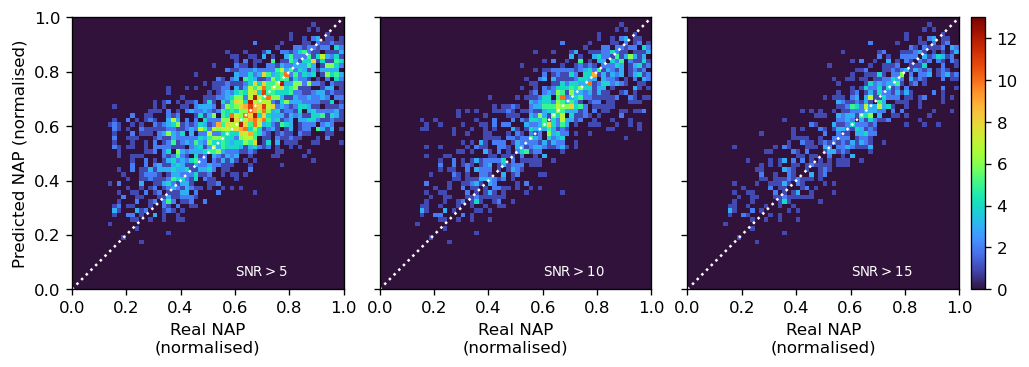}%
    }
    \\
    \subfloat{%
    \hspace*{-7mm}\includegraphics[clip,scale=0.41]{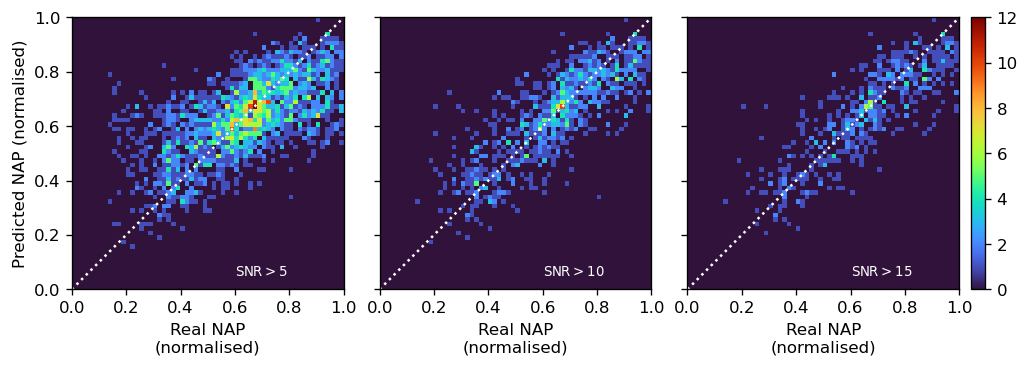}%
    }
    \\
    \subfloat{%
    \hspace*{-7mm}\includegraphics[clip,scale=0.41]{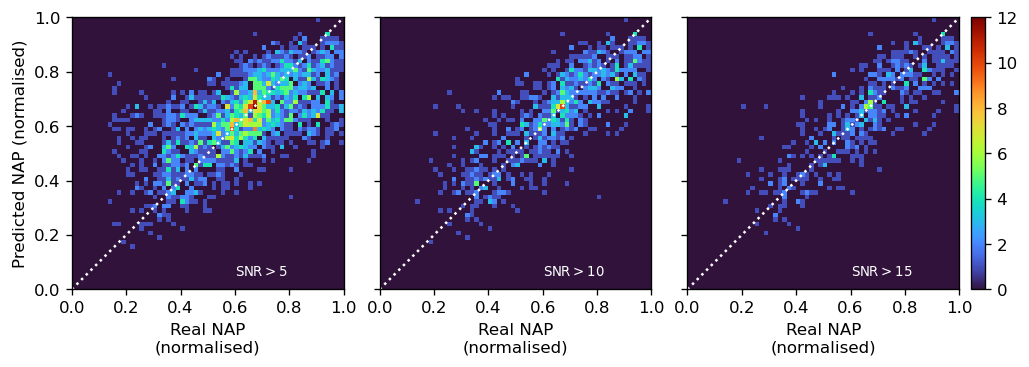}%
    }
    \\
\caption{Same as \cref{fig:multipleapprox_dL} but showing the
calibration results for the network antenna power.}
\label{fig:multipleapprox_NAP}
\end{figure}

In~\cref{fig:multipleapprox_NAP} we show the performance of our model in the regression of the network antenna power (NAP) parameter. Again, as for the luminosity distance, all three approximants show roughly the same behaviour. Also note that almost no events with NAP<0.2 are present, which is to be expected due to the SNR requirements. For the SNR>5 threshold (left column) there seems to be two separate populations, one that broadly follows the diagonal line and a second one that roughly follows a horizontal line around the 0.6 mark. As we increase the SNR threshold, this second population fades away and we isolate a population of predictions that nicely follows the diagonal line.

\subsubsection{Chirp Mass Regression \label{sec:chirp_regression}}

\begin{figure}[t]

\centering
    \subfloat{%
    \hspace*{-5mm}\includegraphics[clip,scale=0.41]{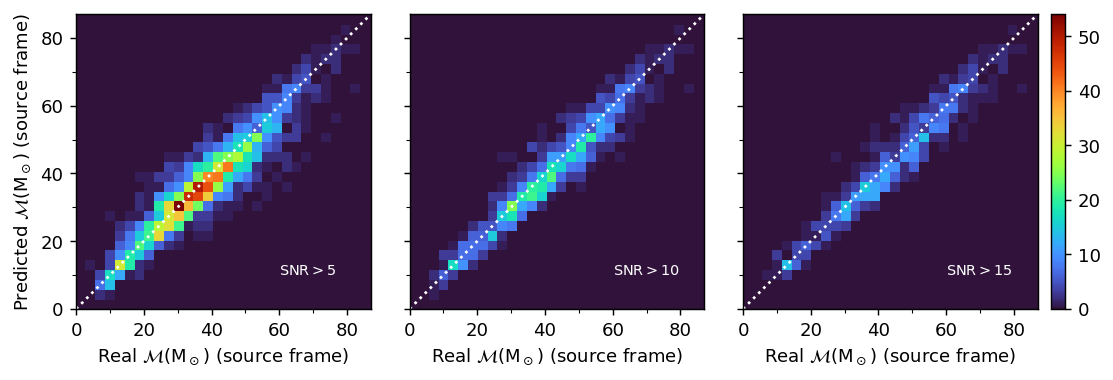}%
    }
    \\
    \subfloat{%
    \hspace*{-5mm}\includegraphics[clip,scale=0.41]{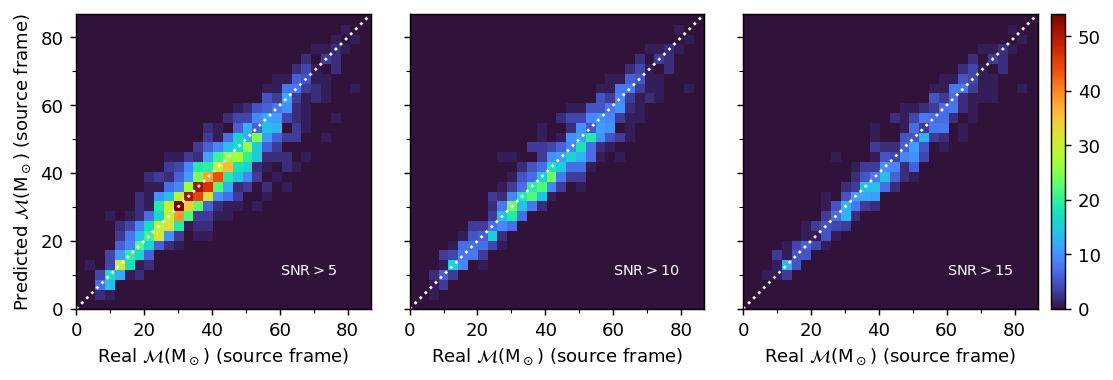}%
    }
    \\
    \subfloat{%
    \hspace*{-5mm}\includegraphics[clip,scale=0.41]{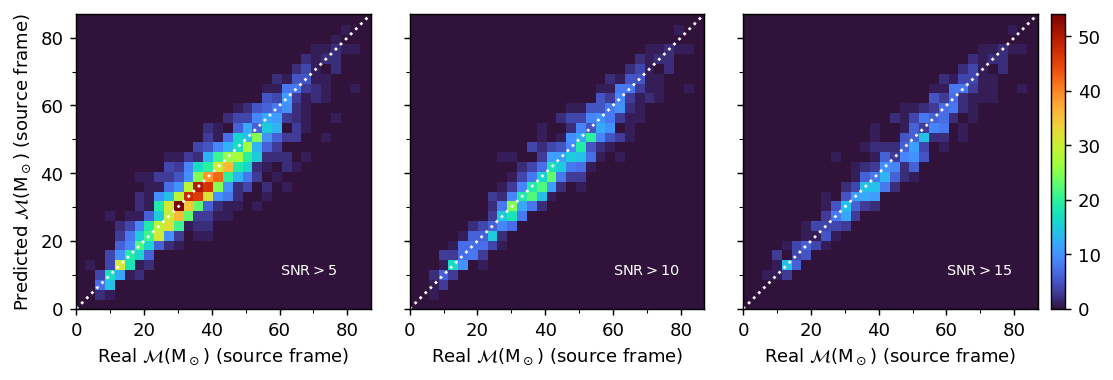}%
    }
    \\
\caption{Same as \cref{fig:multipleapprox_dL} but for
the chirp mass in the source frame.}
\label{fig:multipleapprox_M;}
\end{figure}

\cref{fig:multipleapprox_M;} shows the behaviour of the predicted source frame chirp masses. It is worth mentioning that the bulk of data points in this figure, as well as in \cref{fig:multipleapprox_dL} and \cref{fig:multipleapprox_NAP}, tend to populated the central region of the distributions.
The predictions we obtain for the chirp masses nicely follow the actual injected values as the events closely cluster along the diagonal lines in the plots. Of all variables we employ to calibrate our method, the chirp mass is the one that shows the smallest scatter from the ideal results. The distribution for SNR>5 already displays a fairly low error in the predictions and as we increase the SNR threshold the error further shrinks. Again, the results show almost no dependence on the waveform approximant used.

\subsubsection{Effective Inspiral Spin Regression \label{sec:chi_eff_regression}}

\begin{figure}[t]

\centering
    \subfloat{%
    \hspace*{-7mm}\includegraphics[clip,scale=0.41]{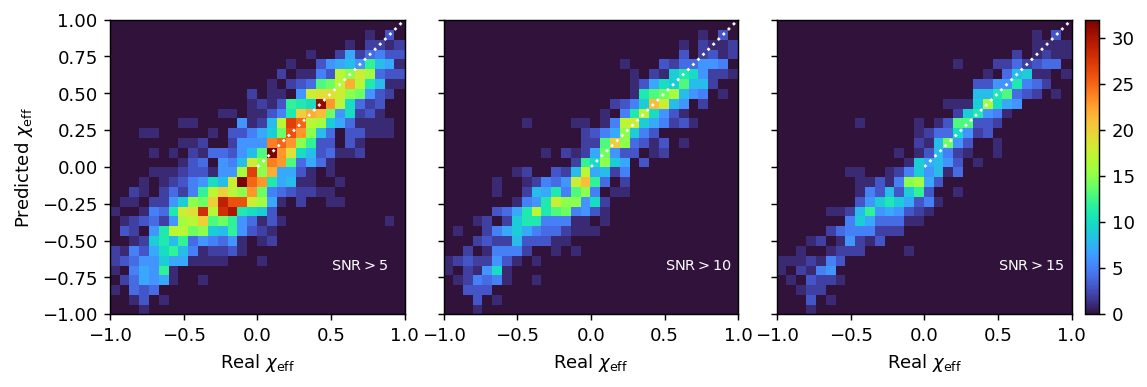}%
    }
    \\
    \subfloat{%
    \hspace*{-7mm}\includegraphics[clip,scale=0.41]{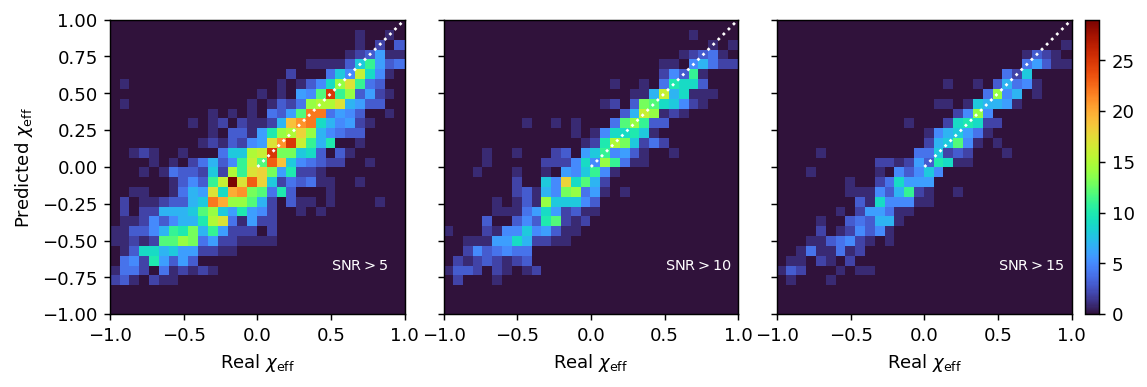}%
    }
    \\
    \subfloat{%
    \hspace*{-7mm}\includegraphics[clip,scale=0.41]{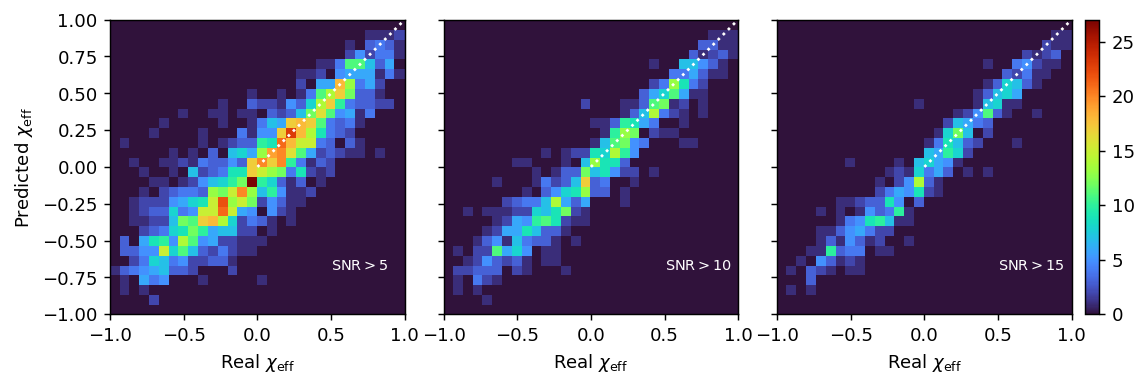}%
    }
    \\
\caption{Same as \cref{fig:multipleapprox_dL} but for the effective inspiral spin.}
\label{fig:multipleapprox_chi;}
\end{figure}

To end the discussion of the calibration of our model we show in \cref{fig:multipleapprox_chi;} the predictions for \chieff\ compared to the real values. In this case we see that all models also follow closely the ideal diagonal line, with the faithfulness of the distribution width increasing as we raise the SNR threshold.

\section{Analysis of real GW detections} \label{sec:comparison}

Having calibrated our method with phenomenological waveforms we turn now to assess its performance with actual GW detections. To this end we initially only selected the BBH detections published in the first GW transient catalog from the LVC, GWTC-1~\cite{GWTC-1}. However, since the GWTC-2 catalog from O3a (comprising the first six months of O3) has recently become publicly available~\cite{GWTC-2} we decided to also apply our methods to the new BBH events reported in the second catalog, for the sake of completeness. It should however be stressed that, as the sensitivity of the detectors improved significantly in the O3 run when compared to O2, the relationship between signal and detector noise differs from the one in our training sets. Therefore, we should a priori not expect optimal results with the current training for the GWTC-2 events. 

\subsection{Classifier}


\begin{table*}
    \def\arraystretch{1.4}
    \setlength\tabcolsep{3pt}
    \centering
\begin{tabular}{|c|c||c|c||c|c|c|c|}
\hline 
\multicolumn{2}{|c||}{GWTC-1 Confident} & \multicolumn{2}{c||}{GWTC-1 Marginal}  & \multicolumn{4}{c|}{GWTC-2}\tabularnewline
\hline 
Event & Score & Event & Score & Event & Score & Event & Score\tabularnewline
\hline 
GW170814 & 1.00 & MC151116 & 0.73 & GW190521 & 1.00 & GW190708\_232457 & 0.98\tabularnewline
GW150914 & 1.00 & MC161217 & 0.72 & GW190602\_175927 & 1.00 & GW190909\_114149 & 0.97\tabularnewline
GW170823 & 1.00 & MC170705 & 0.51 & GW190424\_180648 & 1.00 & GW190514\_065416 & 0.96\tabularnewline
GW170104 & 1.00 & MC170630 & 0.49 & GW190620\_030421 & 1.00 & GW190814 & 0.95\tabularnewline
GW170729 & 0.99 & MC170219 & 0.45 & GW190503\_185404 & 1.00 & GW190521\_074359 & 0.95\tabularnewline
GW170809 & 0.97 & MC161202 & 0.40 & GW190727\_060333 & 1.00 & GW190731\_140936 & 0.92\tabularnewline
GW151012 & 0.96 & MC170423 & 0.35 & GW190929\_012149 & 1.00 & GW190513\_205428 & 0.92\tabularnewline
GW170608 & 0.92 & MC170208 & 0.33 & GW190915\_235702 & 1.00 & GW190421\_213856 & 0.87\tabularnewline
GW170818 & 0.88 & MC170720 & 0.30 & GW190630\_185205 & 1.00 & GW190412 & 0.81\tabularnewline
GW151226 & 0.87 & MC151012A & 0.26 & GW190519\_153544 & 1.00 & GW190728\_064510 & 0.77\tabularnewline
- & - & MC151008 & 0.20 & GW190706\_222641 & 1.00 & GW190719\_215514 & 0.76\tabularnewline
- & - & MC170405 & 0.14 & GW190413\_134308 & 1.00 & GW190803\_022701 & 0.66\tabularnewline
- & - & MC170616 & 0.12 & GW190701\_203306 & 1.00 & GW190930\_133541 & 0.58\tabularnewline
- & - & MC170412 & 0.09 & GW190517\_055101 & 1.00 & GW190828\_065509 & 0.56\tabularnewline
\cline{7-8} \cline{8-8} 
- & - & - & - & GW190408\_181802 & 1.00 & GW190924\_021846 & 0.40\tabularnewline
- & - & - & - & GW190910\_112807 & 1.00 & GW190707\_093326 & 0.35\tabularnewline
- & - & - & - & GW190828\_063405 & 0.99 & GW190720\_000836 & 0.16\tabularnewline
- & - & - & - & GW190413\_052954 & 0.99 & - & -\tabularnewline
- & - & - & - & GW190512\_180714 & 0.98 & - & -\tabularnewline
- & - & - & - & GW190527\_092055 & 0.98 & - & -\tabularnewline
\hline 
\end{tabular}

    \caption{Classifier scores for GWTC-1 marginal detections (left), GWTC-1 confident detections (middle) and GWTC-2 detections (right).}
    \label{tab: ClasScores}
\end{table*}

To analyse the real GW events we produce RGB spectrograms using publicly available data for all GWTC-1 and GWTC-2 BBH events, combining the data from Hanford, Livingston and Virgo. We leave out the binary neutron star events (GW170817, GW190425 and GW190426) as those cases are not present in our datasets and thus we should not expect the network to recognize them. However, we include GW190814 despite the fact that it involves a $\sim2.6M_\odot$ compact object, since its precise nature remains undetermined~\cite{GW190814}. In addition to the confident detections from GWTC-1 and GWTC-2, we also analyze the marginal subthreshold triggers for GWTC-1. 

The results of our classifier are presented in Table \ref{tab: ClasScores}. All confident detections reported by the LVC for GWTC-1 are corroborated by our classifier: GW150914, GW170104, GW170814 and GW170823 are all given the score of 1.00, the highest value possible. From the remaining events, the lowest score is 0.87. When we analyse the marginal detections from GWTC-1 we obtain, as expected, much lower scores across the board. Under the standard threshold for detection, which assumes a score of 0.50 or higher, only three events are classified as signal. These are MC151116, MC161217 and MC170705, with scores of 0.73, 0.72 and 0.51 respectively. The first two in particular may deserve a more careful analysis in the future, but this stays outside the scope of this paper. 
Keeping in mind that the classifier is not optimized for the O3a run, which has a significantly lower noise floor, we look also at the new BBH events of the GWTC-2 catalog. Despite the lack of any optimization, we find that 34 out of 37 events are given a score above our threshold for detection, and from these, 27 (31) events are given a score above 0.90 (0.70). The highest possible score is obtained for a subset of 16 events.

As an aside remark, it is interesting to note that we have also obtained high scores for signals proposed in alternative GW catalogs~\cite{1-OGC, GW151216}. As an example, our method yields a score of 0.75 for GW151216, proposed in~\cite{GW151216}. 

\begin{figure}
\begin{center}
\begin{tabular}{c}
\hspace*{-4mm}\epsfig{file=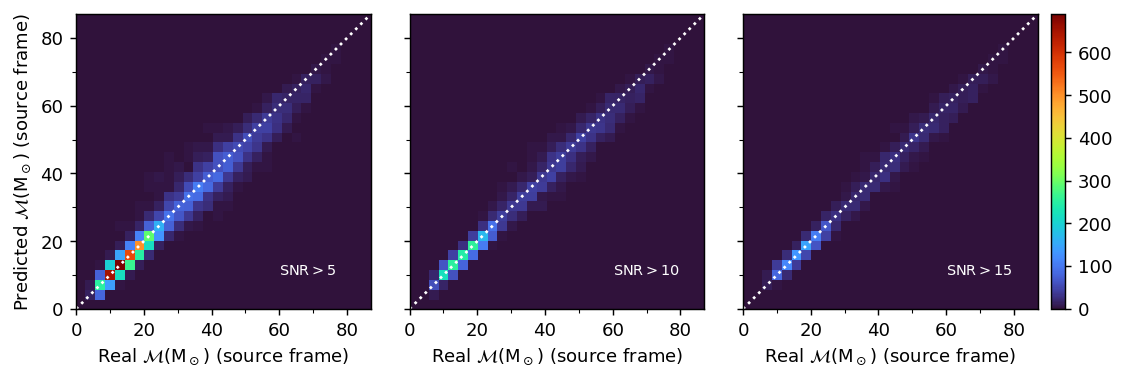,scale=0.41,clip} \\
\end{tabular}
\caption{Calibration results for the chirp mass in the source frame, using the low-mass dataset in addition to the {\tt SEOBNRv4HM\_ROM} dataset, and for different SNR thresholds.}
\label{fig:lowmass_cal_M;}
\end{center}
\end{figure}

\begin{figure*}

\centering
    \subfloat[]{%
    \hspace*{-4mm}\includegraphics[height=7cm,keepaspectratio]{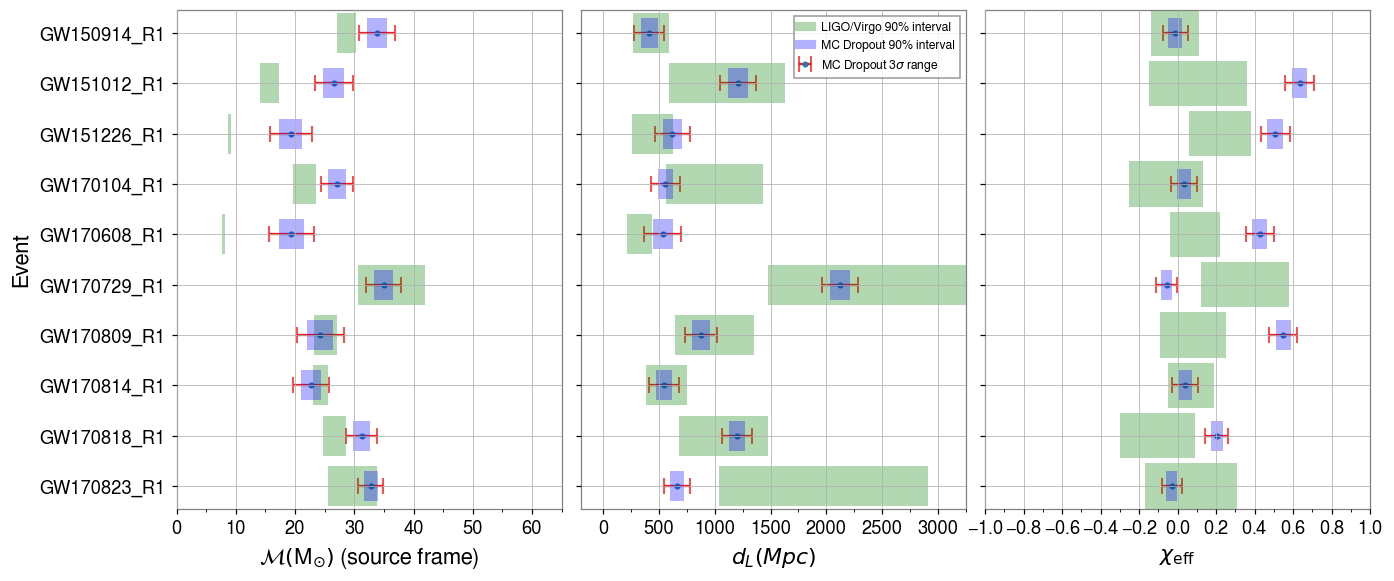}%
    
    }\\
    
    \subfloat[]{
    \centering
    \hspace*{-4mm}\includegraphics[height=7cm,keepaspectratio]{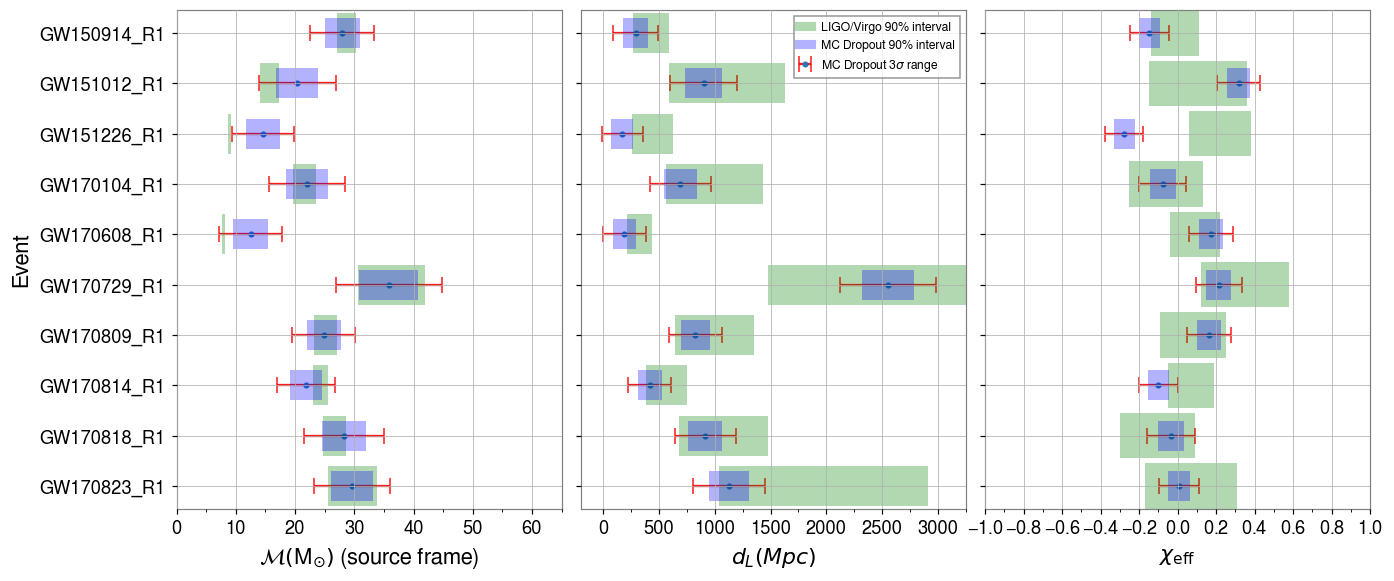}%
    }\\
    \caption{\label{fig:pi_gwtc1}Predictions of the DL network for the chirp mass (left), luminosity distance (middle) and effective inspiral spin (left). The top panels (a) do not include the low-mass correction while the bottom panels (b) do.}
\end{figure*}

\subsection{Parameter Inference on GWTC-1}
\label{sec:regression}

The regression on the BBH events in the LVC catalogs was performed twice. First, with three different networks trained on the three regression datasets, we used MC dropout to pass the spectrogram corresponding to each event to each network, 500 times. We then calculated the mean and standard deviation for the networks' predictions. We realized that with these datasets, low $\mathcal{M}$ events were underrepresented. Therefore, we generated a new dataset using the {\tt SEOBNRv4HM\_ROM} approximant, which was concatenated with the corresponding original dataset, for a total of 31499 items. A new network was trained on this dataset with a 70/30 train/validation split and new calibration plots were produced. In~\cref{fig:lowmass_cal_M;} we show, as an example, the ones corresponding to the chirp masses, for different SNR thresholds. The spectrograms of the catalogued events were then passed to this new network, 1500 times for each event.

For completeness, we show the inference results for both cases i.e.~with and without the contribution of the low mass datasets, in order to highlight the importance of appropriately covering the relevant parameter space. It is worth mentioning  that the results for both cases mostly coincide for the high mass events, which reinforces the consistency of the method. The comparison between our predictions and the values published by the LVC are shown in \cref{fig:pi_gwtc1}.

\subsubsection{Chirp Mass}

In the leftmost panel of~\cref{fig:pi_gwtc1} we show the combined results for our three approximants for the chirp mass $\mathcal{M}$ of the GWTC-1 confident BBH detections. The red error bars enclose the MC dropout 3$\sigma$ range.
The results in the top panel are obtained without including the low-mass distribution while those in the bottom panel do include it. For the former we find that the published 90\% confidence intervals lie outside our predicted $3\sigma$ range for 6 events (without systematic uncertainties taken into account). The most significant discrepancies occur for the GW151226 and GW170608 events where the network seems reluctant to predict low-mass values. We expect this to be related to the under-representation of low chirp mass events in the training set. In fact, when the network is trained with the addition of the low-mass dataset, shown in the bottom panel of Fig.~\ref{fig:pi_gwtc1}, we do indeed observe a considerable improvement. Most of the predictions are now compatible with published data with an uncertainty up to three standard deviations. The one exception is GW151226, where the lower bound on the chirp mass is slightly overestimated by the network when compared to Advanced LIGO results.

\subsubsection{Luminosity Distance}\label{subsec:Lumi}

The top, middle panel of \cref{fig:pi_gwtc1} displays the combined results for the three approximants without the low-mass distribution, for $d_L$ of the GWTC-1 confident BBH detections. For the luminosity distance most of our predictions, except for GW170823, are compatible with the LIGO/Virgo predictions up to a network uncertainty of $3\sigma$. However, when the network is trained with the addition of the low-mass dataset, as displayed in the bottom, middle panel of \cref{fig:pi_gwtc1}, all of our  predictions become compatible with the LVC values.

\subsubsection{Effective inspiral spin}

When infering the effective inspiral spin \chieff\, using the combined results for the three approximants and without including the low-mass distribution, we find six events with a significant disagreement with published results (in the same sense as discussed previously for the chirp mass) as can be seen in the rightmost, top panel of \cref{fig:pi_gwtc1}. As shown in the corresponding bottom panel of the same figure, when the network is trained with the addition of the low-mass dataset, our results improve. All events, except for GW151226, show compatibility between the LVC 90\% confidence interval and the MC dropout $3\sigma$ uncertainty.

\subsection{Parameter Inference on GWTC-2}

\begin{figure*}[t]
    \centering
    \hspace*{-7mm}
    \includegraphics[clip,scale=0.5]{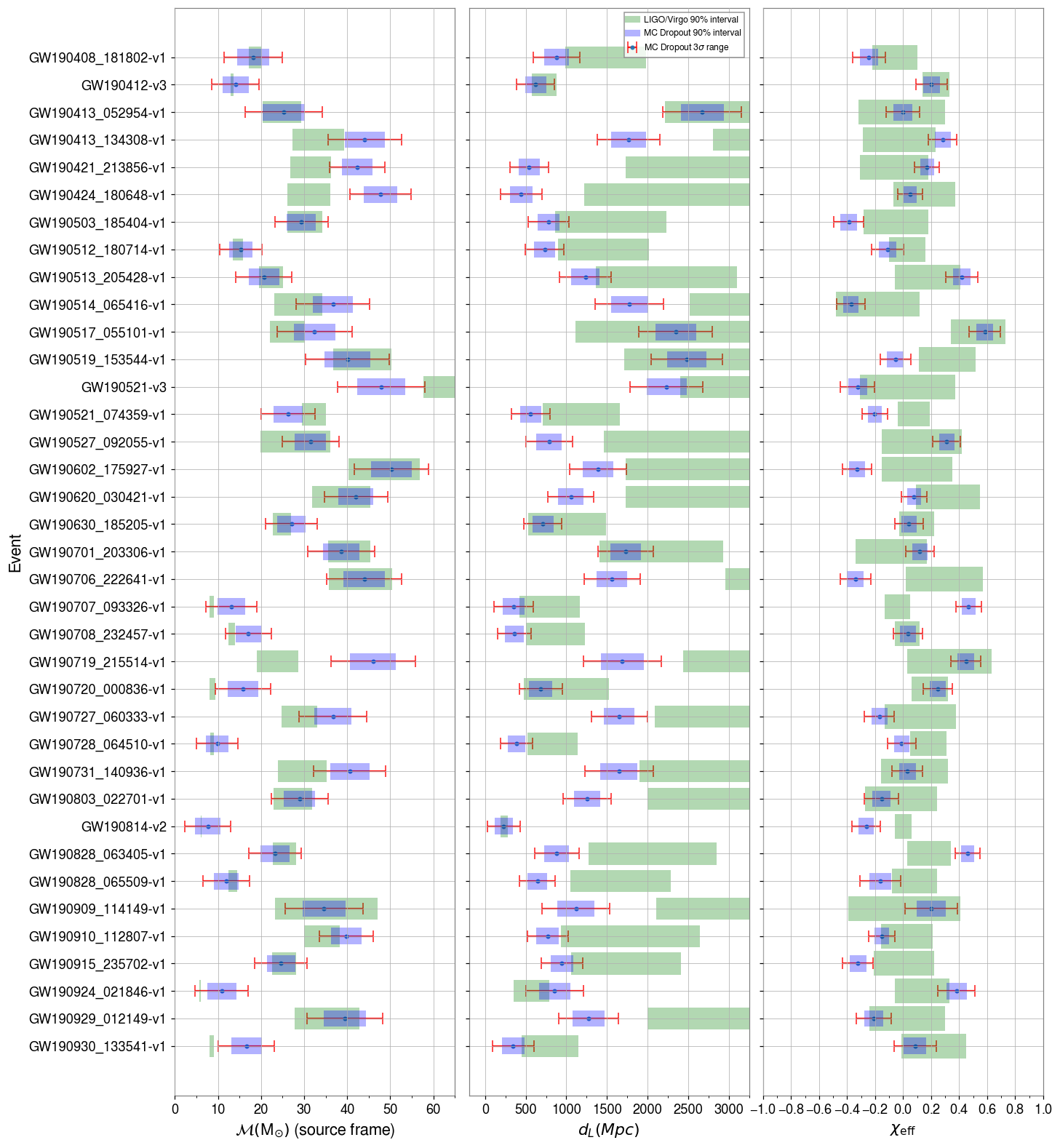}%
    
\caption{\label{fig:pi_gwtc2} Best-performing DL network's prediction for the chirp mass (left), luminosity distance (middle) and effective inspiral spin (left) for the BBH events from GWTC-2.}
\end{figure*}

Finally, we discuss our inference results for the BBH events of GWTC-2. Those are plotted 
in \cref{fig:pi_gwtc2} for our best-performing network only, that is, the one trained with the low-mass corrected dataset. As before, we show the parameter inference on the chirp mass, luminosity distance and effective inspiral spin. Without any further optimization of our network, we find that regarding the chirp mass (left panel), the published 90\% confidence intervals are compatible with the MC dropout $3\sigma$ uncertainty for roughly 90\% of the events. While the behaviour for the \chieff\ predictions tends to also show a reasonable agreement (right panel), the predictions for $d_L$ (middle panel) show a bias towards lower values. We attribute this effect to the usage of O2 noise for the training of the network, as alluded to at the beginning of this section. This conjecture will be analyzed in future work.

\section{Conclusions} \label{sec:conclusions}

In this work we have introduced Deep Learning (DL) methods to study gravitational waves from BBH mergers, using spectrograms created from Advanced LIGO and Advanced Virgo  open data. By combining data from each of the detectors in the Advanced LIGO and Advanced Virgo network using color channels of RGB images we have shown that the classification procedure improves when compared to a single detector case. For black holes of varying mass and zero spin, we have trained a residual network classifier on 2000~Mpc data, obtaining a precision of 0.9 and an accuracy of 0.82. This residual network has been applied on LIGO/Virgo detections and we have corroborated all confident results with high scores, while an analysis of marginal triggers from the O1 and O2 runs has identified 3 cases (MC151116, MC161217 and MC170705) as GW signals, rejecting all others. For GWTC-2 events, despite the lack of any optimization, we have found that 34 out of 37 events pass the threshold for detection.

For the case of black holes of varying mass, spin and sky position, with varying distance, we have trained a cross-residual network to perform parameter estimations on GW spectrogram data from BBH mergers. Using MC dropout we have obtained a natural estimation of the uncertainty of our predictions. We have shown that, at a fledgling level of development, it is possible to successfully perform parameter inference on the distance, chirp mass, antenna power (functioning as a proxy for sky position) and the effective inspiral spin \chieff{}. The success at resolving this last parameter, especially at high SNR values, shows that our method is sensitive to contributions to the Post-Newtonian expansion of the binary system GW radiation up to order 1.5, as this is the first order where a spin-orbit coupling term is observed. Applying this network to spectrogram data from GWTC-1 BBH events, we have found a remarkable agreement with the results published by the LVC in the case of $d_L$ estimations. Most of our chirp mass and effective spin estimations are also compatible with the published 90\% confidence intervals up to an MC dropout uncertainty of $3\sigma$, with the exception of GW151226. For GWTC-2 events, again without any optimization, we have found that the published 90\% confidence intervals for the chirp mass are compatible with our prediction up to $3\sigma$, for 33 out of 37 BBH events. The behaviour for the \chieff\ predictions tends to show a reasonable agreement, similar to that of the chirp mass. The predictions for $d_L$ tend to be underestimated, which we suspect is related with the training of the network with injection on O2 noise, which has different characteristics when compared with the O3 run.

It is important to stress that we have not carried out a thorough search of network architectures in this work. Going forward, optimizing the architecture, as well as exploring other implementations of Bayesian neural networks, could provide further improvements on our results. Other physical effects of BBH mergers, such as orbital plane precession or eccentricity, may also be explored. Lastly, higher resolution spectrograms, as well as higher colour depth, could in theory be used to increase the sensitivity to smaller effects as well as the predictive power of our tool.

\section*{Acknowledgements}
We thank Nicol\'as Sanchis-Gual for very fruitful discussions that allowed to setup the research team involved in this investigation. This work was supported by the Spanish Agencia Estatal de Investigaci\'on  (PGC2018-095984-B-I00), by the Generalitat Valenciana (PROMETEO/2019/071), by  the European Union’s Horizon 2020 research and innovation (RISE) programme    (H2020-MSCA-RISE-2017 Grant No. FunFiCO-777740) and by the Portuguese Foundation for Science and Technology (FCT), project CERN/FIS-PAR/0029/2019. APM and FFF are supported by the FCT project PTDC/FIS-PAR/31000/2017 and by the Center for Research and Development in Mathematics and Applications (CIDMA) through FCT, references UIDB/04106/2020 and UIDP/04106/2020. APM is also supported by the projects CERN/FIS-PAR/0027/2019, CERN/FISPAR/0002/2017 and by national funds (OE), through FCT, I.P., in the scope of the framework contract foreseen in the numbers 4, 5 and 6 of the article 23, of the Decree-Law 57/2016, of August 29, changed by Law 57/2017, of July 19. 

\bibliographystyle{apsrev}
\bibliography{references}

\newpage


\end{document}